\documentclass[12pt,preprint]{aastex}

\usepackage{natbib}
\usepackage{graphicx}

\newcommand{\psr}{PSR~J2021+3651}
\newcommand{\pwn}{G75.2+0.1}
\newcommand{\EG}{{EGRET}}
\newcommand{\gr}{$\gamma$-ray}
\newcommand{\errors}[3]{#1^{+#2}_{-#3}}

\shorttitle{PSR J2021+3651}
\shortauthors{Hessels et al.}

\begin{document}

\title{Observations of PSR~J2021+3651 and its X-ray Pulsar Wind Nebula G75.2+0.1}

\author{J.~W.~T. Hessels, M.~S.~E. Roberts\altaffilmark{1,2}, S.~M. Ransom\altaffilmark{1}, V.~M. Kaspi,}
\affil{Department of Physics, McGill University, Montreal, H3A-2T8, Canada; hessels@physics.mcgill.ca}
\author{R.~W. Romani, C.-Y. Ng,}
\affil{Department of Physics, Stanford University, Stanford, CA 94305}
\author{P.~C.~C. Freire}
\affil{NAIC, Arecibo Observatory, HC03 Box 53995, PR 00612}
\and
\author{B.~M. Gaensler}
\affil{Harvard-Smithsonian Center for Astrophysics, 60 Garden Street MS-6, Cambridge, MA 02138}

\altaffiltext{1}{Also Center for Space Research, M.I.T, Cambridge, MA 02139}
\altaffiltext{2}{Eureka Scientific, 2452 Delmer St. Suite 100, Oakland, CA 94602-3017}

\begin{abstract}
We present results from X-ray and radio observations of the recently discovered young Vela-like pulsar \psr , which is coincident with the \EG\ \gr\ source GeV 2020+3658.  A 19.0-ks {\it Chandra} ACIS-S observation has revealed a $\sim$20$^{\prime \prime} \times 10^{\prime \prime}$ pulsar wind nebula that is reminiscent of the equatorial tori seen around some young pulsars, along with thermal emission from an embedded point source ($kT_{\infty} = 0.15 \pm 0.02$\,keV).  We name the nebula \pwn .  Its spectrum is well fit by an absorbed power-law model with photon index $\Gamma = \errors{1.7}{0.3}{0.2}$, hydrogen column density n$_{\rm H} = (\errors{7.8}{1.7}{1.4}) \times 10^{21} $\,cm$^{-2}$, and an unabsorbed 0.3$-$10.0\,keV flux of $(\errors{1.9}{0.1}{0.3}) \times 10^{-12}$ erg cm$^{-2}$ s$^{-1}$.  We have spatially fit \pwn\ with a model that assumes a toroidal morphology, and from this we infer that the torus is highly inclined ($83^{\circ} \pm 1^{\circ}$) to the line of sight.  A 20.8-ks {\it Chandra} observation in continuous-clocking mode reveals a possible pulse detection, with a pulsed fraction of $\sim$37\% and an H-test probability of occuring by chance of $1.2 \times 10^{-4}$.  Timing observations with the Arecibo radio telescope spanning two years show that \psr\ glitched sometime between MJDs~52616 and 52645 with parameters ${\Delta \nu}/{\nu} = (2.587 \pm 0.002) \times 10^{-6}$ and ${\Delta \dot{\nu}}/{\dot{\nu}} = (6.2 \pm 0.3) \times 10^{-3}$, similar to those of the largest glitches observed in the Vela pulsar.  \psr\ is heavily scattered (${\tau}_{sc} = 17.7 \pm 0.9$\,ms at 1\,GHz) and exhibits a significant amount of timing noise.
\end{abstract}


\section{Introduction}

Pulsar wind nebulae (PWNe) allow us to study the basic energetic processes of young, rotation-powered pulsars.  It is still unclear how the rotational kinetic energy of these objects is converted into a relativistic particle outflow.  The {\it Chandra X-ray Observatory}'s unprecedented spatial resolution has revealed that the winds from young, energetic pulsars can lead to highly structured PWNe.  In some cases these PWNe take the form of an equatorial outflow with polar jets, where the best example of this is the Crab nebula \citep{wht+00}.  These torus plus jet nebulae, of which there are roughly a half-dozen examples, are particularly interesting because a) they allow us to study the interaction of the PWN with its surrounding environment (the location of the bright torus is generally believed to mark the inner termination shock of the pulsar wind) and b) the well-defined morphology of the PWN can constrain the geometry of the pulsar.  This is crucial for testing models of high-energy pulsations, such as the outer-gap model \citep[e.g.][]{rom96} and the polar-cap model \citep[e.g.][]{dh96}.

\psr\ (spin period $P \sim$103.7\,ms, dispersion measure
 DM $\sim$369\,pc cm$^{-3}$, and flux density at 1400\,MHz $S_{1400} \sim$100\,$\mu$Jy) was discovered by \citet{rhr+02} with the 305-m Arecibo radio telescope during a targeted search for radio pulsations from X-ray sources proposed as counterparts to unidentified \EG\ \gr\ 
sources \citep*{rrk01}.  Timing observations of \psr\ revealed that although it is dim, it is young and energetic (characteristic age ${\tau}_c \equiv P/2\dot{P} = 17$\,kyr, $\dot{E} \equiv 4{\pi}^2 I \dot{P}/P^3 = 3.6 \times 10^{36}$\,ergs s$^{-1}$, assuming moment of inertia $I = 10^{45}$ g cm$^{-2}$), and a likely counterpart to the {\it ASCA} X-ray source AX J2021.1+3651 and the \EG\ \gr\ source GeV J2020+3658.  This discovery bolsters the idea that many of the Galactic unidentified \EG\ sources \citep[see][]{hbb+99} are young energetic pulsars that were not detected in large-scale radio pulsar surveys. 

While the spin properties of \psr\ are very similar to those of the known \gr\ pulsars Vela (PSR~B0833$-$45, $P = 89.3$\,ms, $\dot{E} = 6.9 \times 10^{36}$\,ergs s$^{-1}$) and PSR~B1706$-$44 ($P = 102.5$\,ms, $\dot{E} = 3.4 \times 10^{36}$\,ergs s$^{-1}$), the 12\,kpc distance inferred from its DM (which places it on the far edge of the outer spiral arm) using the NE2001 electron model of the Galaxy \citep{cl02} would make it much more \gr\ efficient than either Vela or PSR~B1706$-$44.  In fact, it would have to be even more \gr\ efficient than the much-less energetic \gr\ pulsars Geminga and PSR~B1055$-$52 (assuming the NE2001 DM distance to PSR~B1055$-$52 of 0.72\,kpc).  This is contrary to the observation that \gr\ efficiency ${\eta}_{\gamma} \equiv L_{\gamma}/\dot{E}$ increases with age \citep{thom99} and the theoretical expectation that it is inversely proportional to the square-root of the spin-down luminosity \citep[${\eta}_{\gamma} \propto {\dot{E}}^{-1/2}$, e.g.][]{zh00a}.  However, since DM distances can be off by a factor of a few, \psr\ may be significantly closer (or farther) than its DM implies.  

The similarity of \psr\ to the Vela pulsar, which is surrounded by a highly-structured pulsar wind nebula \citep*[see][]{hgh01}, prompted us to make X-ray observations of the area surrounding \psr.  Only $\sim$30 of the $>$1300 known pulsars in the Galaxy are Vela-like \citep[i.e. with $P \sim 100$\,ms, $\dot{E} \sim 10^{36}-10^{37}$ ergs s$^{-1}$, and characteristic ages $\tau_c \sim$10$-$20\,kyr, see][]{kbm+03}.  Half of these are known to have associated PWN \citep*{krh04}.  Here we present the results of X-ray observations made with the {\it Chandra} ACIS-S detector.  These observations resolve the {\it ASCA} source into a new PWN, which we name \pwn , and a thermally emitting point source, likely the neutron-star surface.  {\it Chandra} observations in continuous-clocking mode show a potential detection of X-ray pulsations from \psr .  We also present radio timing observations made with the Arecibo telescope, which show that \psr\ has glitched and is highly scattered.

\section{X-ray Observations}

\subsection{X-ray Imaging}

\begin{figure}
\plotone{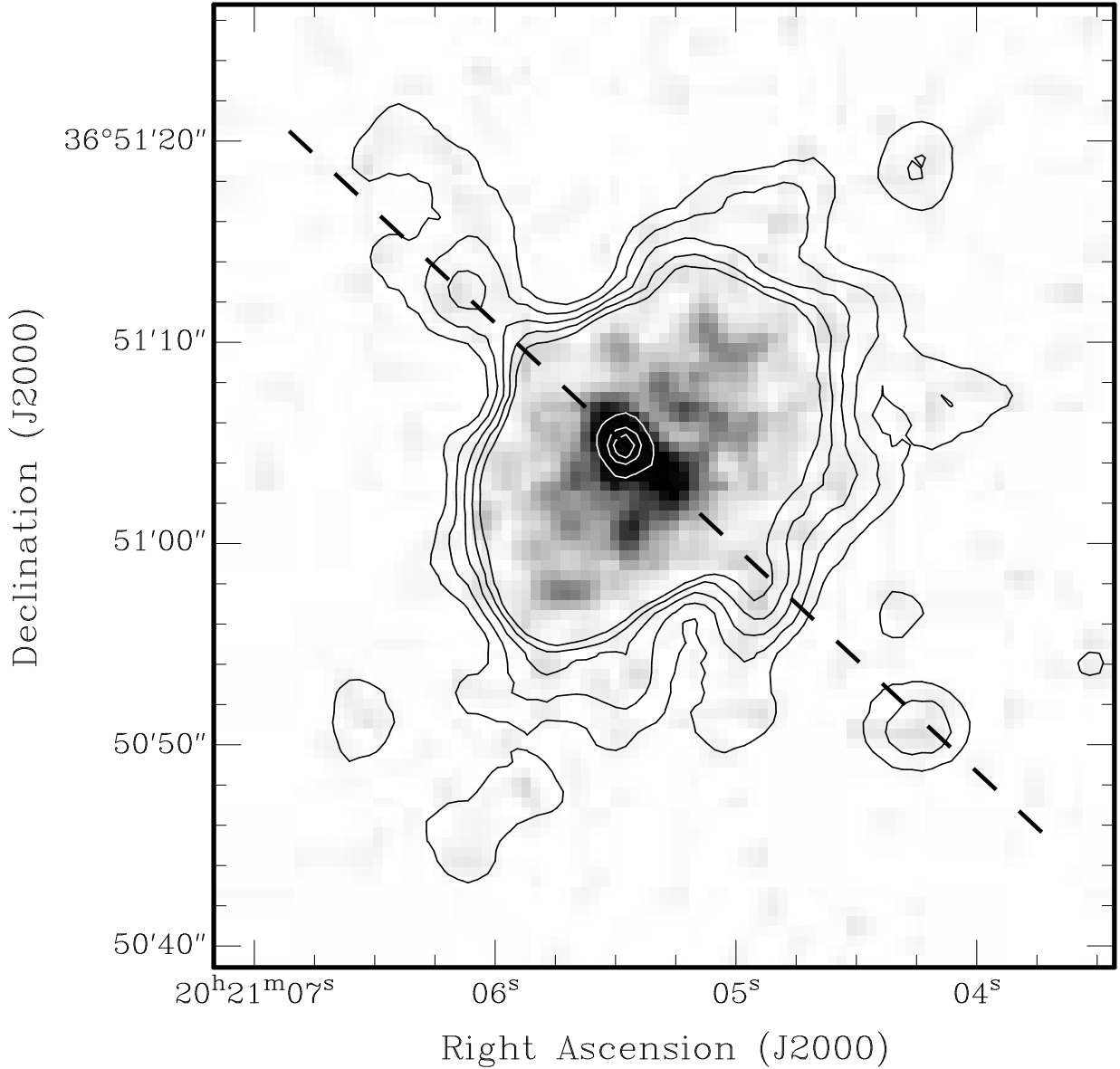}
\caption{ A $19.0$-ks {\it Chandra} ACIS-S image (0.3$-$7.0\,keV) of \psr\ and its inner nebula.  The greyscale shows the image smoothed with a $1.2^{\prime \prime}$ (FWHM) gaussian and scaled to bring out fine structure in the nebula. 
The black contours are the same image overlaid, but smoothed with a $3.5^{\prime \prime}$ gaussian and scaled to bring out faint extended emission including what may possibly be faint jets along the axis of the nebula.  The white contours indicate the point source and are shown to distinguish it from the bright bar running along the axis of the nebula.  The dashed line shows the best-fit position angle $\Psi = 47^{\circ} \pm 1^{\circ}$ of the torus fit (see \S 4.4), which in analogy with the Crab pulsar and nebula likely corresponds to the projected spin axis of the pulsar.  
\label{pwnfig}}
\end{figure}

We made {\it Chandra} observations with 
the ACIS-S detector in VFAINT mode on 2003 February 12 to image \psr\ and its surrounding region.  To reduce CCD photon pile-up from a potentially bright point source, a quarter sub-array was used.  These data were analyzed using CIAO\footnote{see http://cxc.harvard.edu/ciao} version 3.0.1 and CALDB 2.3.  Filtering for good time intervals and accounting for dead-time resulted in a total integration time of 19.0\,ks. 
We subtracted particle background from the image using a 100-ks combination of two ACIS-stowed images from 2002 Sept and 2003 May\footnote{see http://cxc.harvard.edu/contrib/maxim/acisbg} and then corrected for the different exposure over the chip.  The resulting image is shown in Figure~\ref{pwnfig}.

This image reveals a point source embedded in an axisymmetric inner nebula $\sim$20$^{\prime \prime} \times 10^{\prime \prime}$ across, surrounded by fainter diffuse emission.  We refer to this obvious PWN as \pwn .  The point source is at right ascension $20^h21^m 05^s.46$ and declination $+36^{\circ}51^{\prime}04^{\prime \prime}.8$ (J2000).  We estimate a positional error of $\sim$1$^{\prime \prime}$ by comparing the position of another bright X-ray point source on the S3 chip with a catalogued optical counterpart.  Subtracting the nebular background, the point source has a count rate of $0.0094 \pm 0.0009$\,counts s$^{-1}$ and the nebula has a background-subtracted count rate of $0.075 \pm 0.004$\,counts s$^{-1}$ in the 0.3$-$10.0\,keV range.

The greyscale of Figure~\ref{pwnfig} shows the image smoothed with a $1.2^{\prime \prime}$ (FWHM) gaussian to bring out the fine structure of the inner nebula.  The white contours indicate the position of the point source and are shown to distinguish it from the bright bar running along the axis of the nebula.  This bar is the most obvious resolved feature of the nebula.  Also overlaid are black contours of the same image smoothed with a $3.5^{\prime \prime}$ gaussian and scaled to bring out what are possibly faint jets along the axis of the nebula.  We estimate 62$\pm$32 counts may be coming from these putative jets.  The nebula appears symmetric about the minor axis, but the point source is clearly offset from the major axis of the nebula.  The overall morphology is suggestive of the torus plus weak jet morphologies of the Crab and (likely) Vela nebulae \citep{wht+00,hgh01}.  We discuss the geometry of the PWN further in \S 4.4.

\subsection{X-ray Spectroscopy}

\begin{figure}
\plottwo{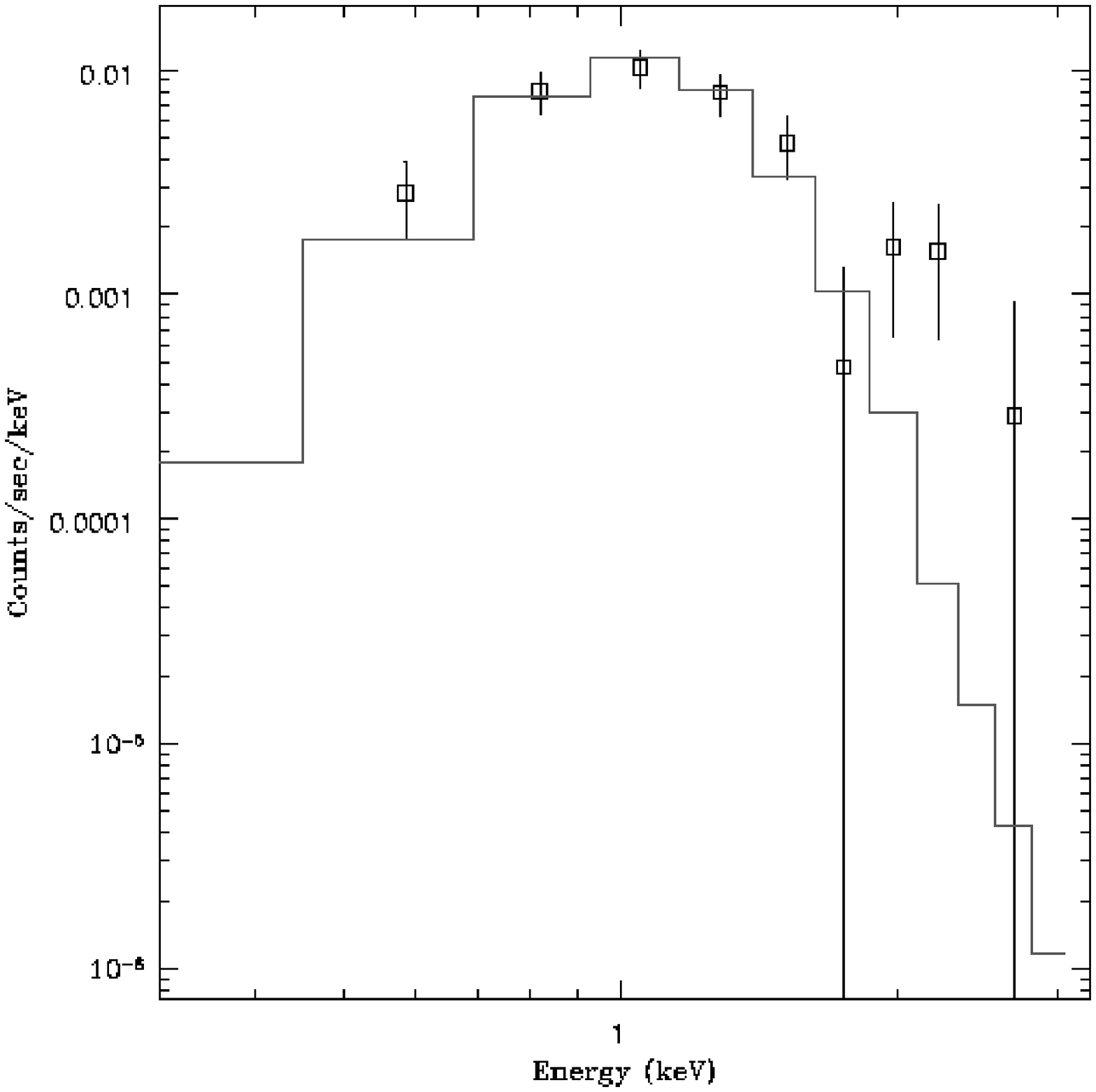}{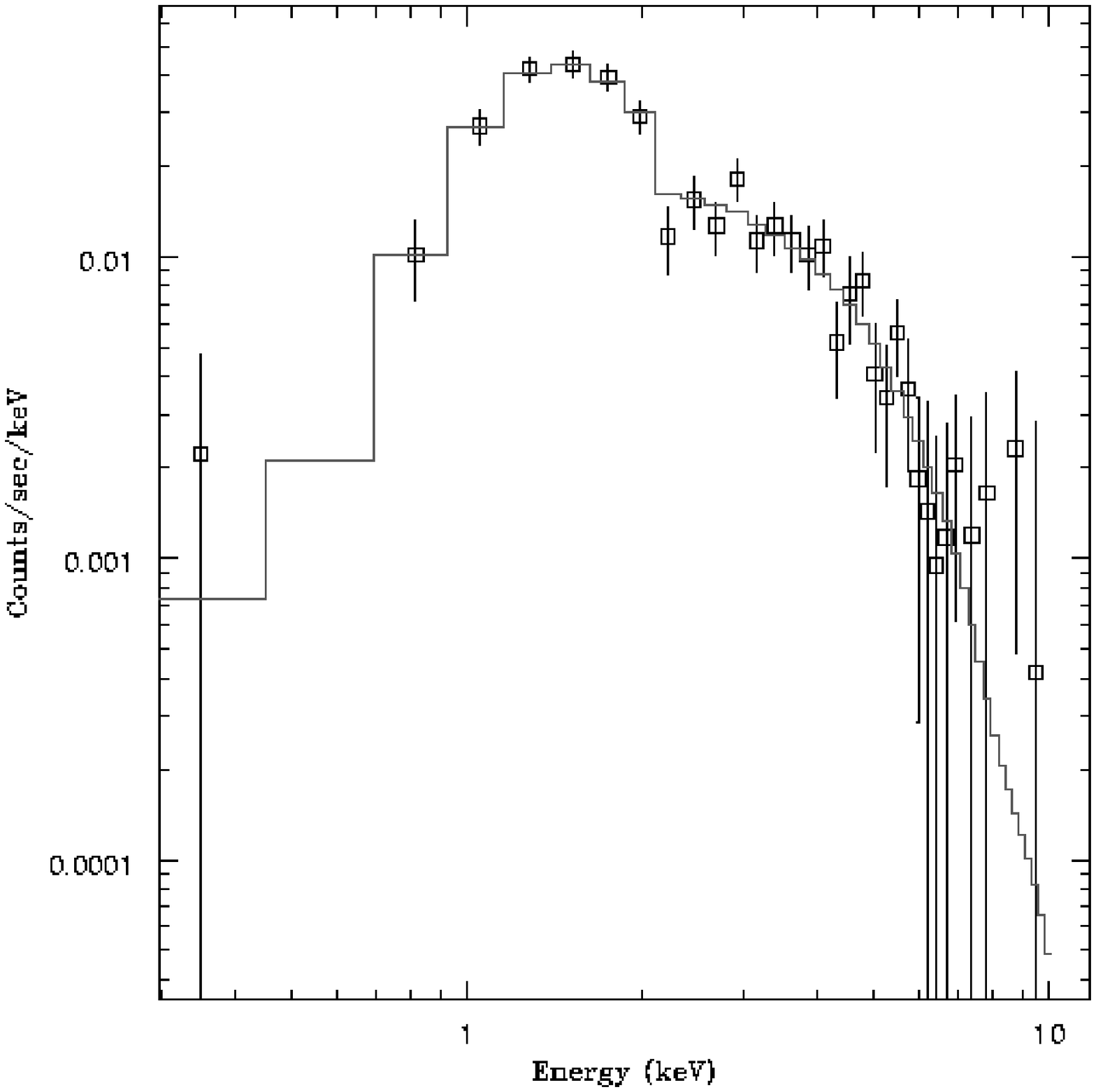}
\caption{ {\it Left}: The background-subtracted spectrum of the point source, 
fit using an absorbed black-body model in {\it Sherpa} (CIAO 3.01).  The best-fit temperature and 90\% confidence interval is
$kT_{\infty} = 0.15 \pm 0.02$\,keV with a 0.3$-$10.0\,keV absorbed flux of $(2.8 \pm 0.2) \times 10^{-14}$ erg cm$^{-2}$ s$^{-1}$.  The n$_{\rm H}$ was fixed at $7.8 \times 10^{21}$~cm$^{-2}$, that found by fitting the nebular spectrum.
{\it Right}: The background-subtracted spectrum of the nebula fit with an absorbed power-law model.  
The best-fit parameters and 90\% confidence intervals 
are n$_{\rm H}=(\errors{7.8}{1.7}{1.4})\times 10^{21}$~cm$^{-2}$, $\Gamma = \errors{1.7}{0.3}{0.2}$, and an absorbed flux of $(1.2 \pm 0.1) \times 10^{-12}$ erg cm$^{-2}$ s$^{-1}$ (0.3$-$10.0\,keV).
\label{specfig}}
\end{figure}

The spectra of the PWN and point source can be measured individually using {\it Chandra's} excellent spatial resolution.  We fit the spectrum of the nebula and point source using CIAO's {\it Sherpa} environment (version 3.0.1) and XSPEC (version 11.2.0).  To isolate the nebula, we used a circular annulus centered on the point source and extending from a radius of 1.6$-$100 pixels ($0.492^{\prime \prime} \times 0.492^{\prime \prime}$ per pixel) and a circular region from an apparently flux-free area of the S3 chip to estimate the background.  We find that the nebula has a hard spectrum (see Figure~\ref{specfig}) that is well fit (reduced ${\chi_\nu}^2 = 1.1$ for 27 degrees of freedom) by an absorbed power-law model with hydrogen column density n$_{\rm H}=(\errors{7.8}{1.7}{1.4})\times 10^{21}$~cm$^{-2}$, photon index $\Gamma = \errors{1.7}{0.3}{0.2}$, and a 0.3$-$10\,keV absorbed flux of $(1.2 \pm 0.1) \times 10^{-12}$ erg cm$^{-2}$ s$^{-1}$ (90\% confidence intervals).  The best-fit black-body model gives reduced ${\chi_\nu}^2 = 2.3$ for 27 degrees of freedom, a comparatively poor fit to the data.

To isolate the point source, we used a circular extraction region of radius 1.6 pixels, and a surrounding annulus of width 1.5 pixels to subtract the nebular background.  Because the energy dependent PSF extends significantly past 1.6 pixels at this location, we corrected for this by using ``mkpsf" to generate model PSFs at several energies between 0.5$-$6\,keV from which we estimated the source minus background enclosed energy fraction.  We then fit a linear function to these points which we used to correct the effective area model used in the spectral fit.
Based on the observed count rate, we estimate a negligible pile-up fraction of $\sim$0.6\% using HEASARC's webPIMMs tool\footnote{see http://heasarc.gsfc.nasa.gov/Tools/w3pimms.html}.  We find that there is very little emission from the point source above 3\,keV.  Its spectrum is adequately fit (reduced ${\chi_\nu}^2 = 1.7$ for 9 degrees of freedom) by an absorbed black-body model, where we have fixed n$_{\rm H}$ to the value obtained by fitting the PWN's spectrum.  The best-fit black-body model and 90\% confidence region is $kT_{\infty}$ $ = 0.15 \pm 0.02$\,keV with an absorbed flux of $(2.8 \pm 0.2) \times 10^{-14}$ erg cm$^{-2}$ s$^{-1}$ (0.3$-$10.0\,keV).  Since $kT_{\infty}$ and n$_{\rm H}$ are highly covariant, we also made a joint fit of an absorbed black body for the pulsar and an absorbed power law for the nebula, tying the n$_{\rm H}$ of both models together.  This gave values and errors that were essentially the same as those from the individual fits.  Since we expect non-thermal emission from the magnetosphere of the pulsar, we also fit a black-body plus power-law model to the point source, fixing the photon index at a typical value of 1.5.  This did not improved the fit (reduced ${\chi_\nu}^2 = 1.8$ for 7 degrees of freedom), but the temperature and flux of the black-body component were left relatively unchanged.  This suggests a maximum non-thermal absorbed flux of $\sim 3 \times 10^{-14}$ erg cm$^{-2}$ s$^{-1}$ (0.3$-$10.0\,keV).

The total flux measured here is a factor of four smaller than seen by {\it ASCA} \citep{rrk01,rhr+02}.  We note that the {\it ASCA} extraction region has a radius of $\sim$4$^{\prime}$, and thus includes areas well outside the $2^{\prime}$-wide field of view of the quarter array used here.  A highly smoothed image of the {\it Chandra} data suggests there may be additional diffuse emission that could extend well beyond the chip.  Furthermore, examination of the {\it ASCA} image suggests there may be additional point sources or structures within the {\it ASCA} extraction region that are not within the {\it Chandra} field of view.
However, despite the flux discrepancy, the derived n$_{\rm H}$, nebular photon index and point source $kT_{\infty}$ are consistent with the {\it ASCA} measurements of the source.

\subsection{X-ray Pulsations}

\begin{figure}
\plotone{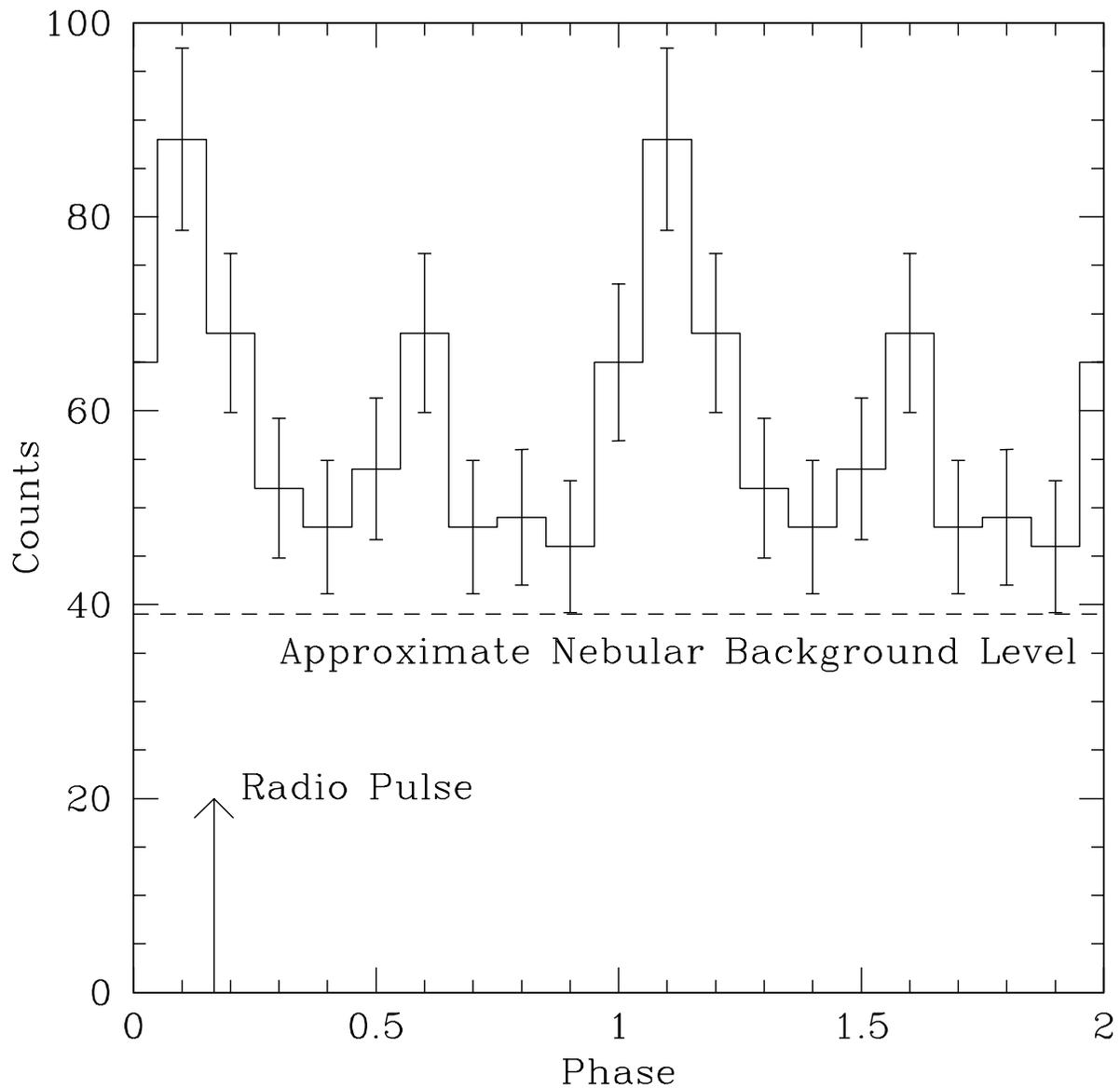}
\caption{ {\it Chandra} continuous clocking data of \psr\ folded using a local radio ephemeris for the pulsar.  Two cycles are plotted for clarity and there are ten bins across the profile.  The phase of the radio pulse is indicated by an arrow.  The dashed line indicates the approximate background level.  Based on this, there are $\sim$69 pulsed counts and we estimate that $\sim$190 counts are from the pulsar.  This gives a pulsed fraction of $\sim$37\%.
\label{Xraypulsefig}}
\end{figure}

To search for X-ray pulsations from \psr , we also obtained {\it Chandra} ACIS-S data in continuous-clocking mode, which provides an effective time resolution of 2.85\,ms by sacrificing one dimension of spatial resolution.  The roll angle of the telescope placed the 1 dimensional image at an angle of $\sim$62$^{\circ}$ (N-E).  The events were filtered for good time intervals and periods of high background, leaving 20.8\,ks of integration time.  We corrected the read-out times to intrumental times of arrival by removing telescope dither, motion of the detector with respect to the telescope, and the average lag between the readout time and the true times of arrival of the photons.  The events were then barycentered using the JPL DE405 ephemeris \citep{stan98}.  Given prior knowledge of the point-source spectrum from our imaging observations, we produced an event list with a 0.5$-$3\,keV energy cut and a 3-pixel-radius extraction region.  We also produced event lists for a number of other energy cuts and a range of extraction regions and then dithered the events by adding random deviates in the range 0$-$2.85\,ms to remove windowing effects.  The event lists were folded at the predicted spin frequency of 9.64094861\,Hz according to the contemporaneous radio timing ephemeris (see \S 3), which revealed a likely periodic signal (see Figure~\ref{Xraypulsefig}).  The signal shows up for a range of extraction region sizes and energy cuts, but has a maximum signal strength for an extraction region of radius 3 pixels with a 0.5$-$3\,keV energy cut.  Performing a variant of the H-test \citep{dsr89}, using {\it unbinned} events, we find an H-test score of $h\sim$11.4, which corresponds to a chance probability of $1.2 \times 10^{-4}$ for a single trial.  This has an equivalent Gaussian sigma of 3.7\,$\sigma$.  Searching to a maximum harmonic of 20, the H-test finds the largest significance when two harmonics are summed.

	Although this signal is not a certain detection of pulsations from \psr , it is intriguing.  If real, these pulsations indicate a pulsed fraction of $\sim$37\%, where we define pulse fraction to be the ratio of the number of counts above the 1-$\sigma$ upper limit on the lowest bin in the light curve (see Figure~\ref{Xraypulsefig}) to the total number of background-subtracted counts (in this case, most of the background is from the nebula).  From the point source count-rate of our imaging observations, we estimate that $\sim$190 of the counts in the light curve are from the pulsar and that the rest are from the nebula and other background.  We discuss the pulse shape in \S 4.2.

\section{Radio Timing}

\begin{figure}
\plotone{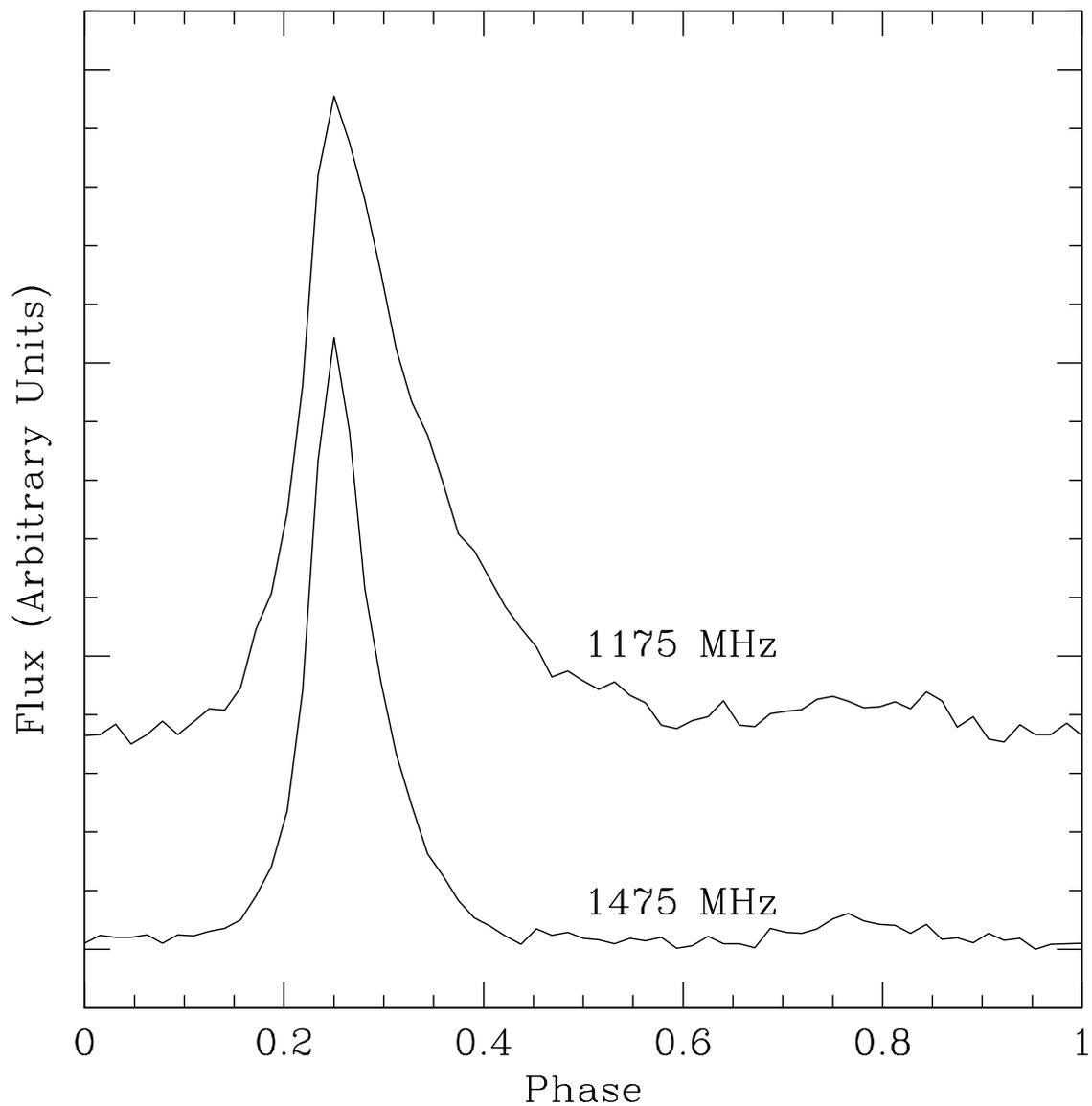}
\caption{Summed pulse profiles of \psr\ at center frequencies of 1175\,MHz (top) and 1475\,MHz (bottom).  The peak fluxes of the two profiles are normalized and the pulse phases aligned at the pulse peak.  By fitting the profiles to a gaussian convolved with an exponential we calculate a scattering time ${\tau}_{sc} = 17.7 \pm 0.9$\,ms at 1\,GHz.  \label{profiles.fig}}
\end{figure}

Since the discovery of \psr\ in 2002 February, we have been making roughly monthly radio timing observations of the pulsar at 20\,cm using the Arecibo radio telescope and multiple Wideband Arecibo Pulsar Processor digital correlator backends (WAPPs, for details see Dowd, Sisk, \& Hagen 2000\nocite{dsh00}).  Each WAPP was configured for 100\,MHz of bandwidth, with 512 lags and 200\,$\mu$s samples.  Whenever possible, we used three WAPP backends, typically centered at 1170, 1420, and 1520\,MHz.  \psr\ is significantly scattered (see Figure~\ref{profiles.fig}).  Fitting the profiles with a gaussian convolved with a one-sided exponential ($e^{-t/{\tau}_{sc}}$) we find a frequency dependence of ${\tau}_{sc}$ that is well fit by a power law of index $-4.2 \pm 0.6$.  This is consistent with a pure Komolgorov spectrum \citep[i.e. ${\tau}_{sc} \propto {\nu}^{-4.4}$, see][]{ric77}.
Using the measured frequency dependence, we find ${\tau}_{sc} = 17.7 \pm 0.9$\,ms at 1\,GHz (90\% confidence).  

Pulse profiles from each epoch were cross-correlated with a representative high signal-to-noise template to generate times of arrival (TOAs).  Using {\tt TEMPO}\footnote{see http://www.pulsar.princeton.edu/tempo} we fit these TOAs to a multi-parameter rotational model of the pulsar.  The position of the pulsar was fixed at the {\it Chandra} X-ray position.
Our timing analysis reveals that a glitch occured between MJDs 52616 and 52645.  We can connect phase across it by fitting an instantaneous jump in frequency and frequency derivative (${\Delta \nu}/{\nu} = 2.587 \pm 0.002 \times 10^{-6}$ and ${\Delta \dot{\nu}}/{\dot{\nu}} = 6.2 \pm 0.3 \times 10^{-3}$), which are comparable to those seen in the largest glitches of the Vela pulsar \citep{dml02}.  There is only marginal evidence for a glitch decay in the data, which is not surprising given that the typical timescale for such a decay is on the order of days to weeks \citep{dml02} and is therefore not resolved by our sparse monitoring.  The pre- and post-glitch rotational ephemerides of \psr\ and the complete set of glitch parameters are presented in Table~1, along with derived quantities.

\begin{figure}
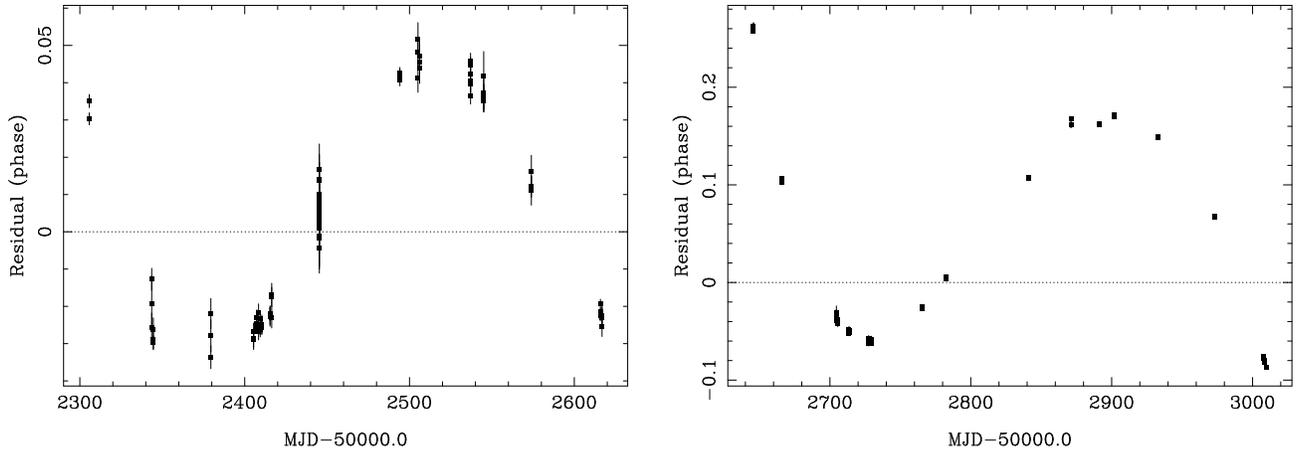

\centerline{\includegraphics[height={0.5\columnwidth}, angle=270]{f5a.eps} \hspace{3mm} \includegraphics[height={0.5\columnwidth}, angle=270]{f5b.eps}}
\caption{Residuals from phase-coherent timing of \psr\ fitting only for period and the first period derivative.  The residuals can be rendered featureless by fitting for a period second derivative.  Alternately, the residuals can be rendered featureless by fitting for position (instead of keeping it fixed at the X-ray position).  However, the fitted positions from the pre- and post-fit ephemerides are inconsistent suggesting that timing noise dominates the residuals, rather than an incorrect position.  {\it Left:} pre-glitch residuals.
{\it Right:} post-glitch residuals.
  \label{timfig}}
\end{figure}

The post-fit timing residuals show significant systematics when only the frequency and first frequency derivative are fit (see Figure~\ref{timfig}).  We find that the residuals can be rendered featureless by including a frequency second and third derivative into the fit.  Alternately, the residuals can be rendered featureless by allowing {\tt TEMPO} to fit for position.  However, the pre-, post-glitch, and {\it Chandra} positions are inconsistent, suggesting that timing noise dominates the residuals rather than an incorrect position.

\section{Discussion}

\subsection{Distance}

As discussed in \citet{rhr+02}, the DM distance of \psr\ is quite large \citep[$\sim$12\, kpc using the NE2001 electron density model of the galaxy, see][]{cl02} and would place \psr\ on the outer edge of the outer spiral arm of the Galaxy.  This large distance leads to a very high inferred \gr\ efficiency (${\eta}_{\gamma} \equiv L_{\gamma}/\dot{E} \sim 0.18$ in the 100\,MeV$-$10\,GeV range, assuming 1-sr beaming).  \citet{cl02} note that the electron density model is seldom off by more than 50\% when predicting distances, except when the line of sight intersects a particularly dense scattering screen.  We find a scattering measure SM = $(1.5 \pm 0.1)~d_{10}^{-5/6}$\,kpc m$^{-20/3}$ towards \psr , where ${\tau}_{sc} \equiv 1.10$\,ms SM$^{6/5} {\nu}^{-22/5} d$ \citep{cl02},  that is two orders of magnitude larger than predicted by the NE2001 model (although \psr\ is in the direction of the Cygnus region and pulsars in that direction are often highly scattered).  This suggests the presence of a clump of scattering material along the line of sight not accounted for by the NE2001 model.  However, small clumps of material that only change the DM by 10$-$20\,pc cm$^{-3}$ ($< 5$\% in the case of \psr) can change the SM by a factor of a hundred.  Hence, the large SM does not by itself imply a significantly closer distance.  In order for the distance to imply a much smaller \gr\ efficiency, a significant source of dispersion ($\sim$200\,pc cm$^{-3}$) unaccounted for by the model is necessary.  Such a DM enhancement could come from an ultra-compact H{\sc ii} region or an OB association.  An examination of the Midcourse Space Experiment 8.3\,$\mu$m IR data (available from the NASA/IPAC Infrared Science Archive) and preliminary radio imaging with VLA data we have obtained (work in preparation) show no evidence for an ultra-compact H{\sc ii} region.  The OB association Cyg 1,8,9 lies in the direction of \psr \citep{yr97,me95}, however there is currently no evidence that this is significantly contributing to the amount of ionized gas along the line of sight or that the pulsar was born in this association.  Since the DM distance places \psr\ at the very outer edge of the Galaxy one would expect our spectral fit to give an n$_{\rm H}$ consistent with the total column density in this direction.  The X-ray absorption found here, n$_{\rm H}=(\errors{7.8}{1.7}{1.4})\times 10^{21}$~cm$^{-2}$, is lower than the total Galactic H{\sc i} column density in the direction of \psr , $1.2 \times 10^{22}$~cm$^{-2}$, estimated using the FTOOL {\it nh}\footnote{see http://heasarc.gsfc.nasa.gov} \citep[which uses the H{\sc i} map of][]{dl90}.    This argues for a distance somewhat closer than the predicted DM distance, but not closer by factor of a few.  Hence a distance of $\sim$8\,kpc is quite plausible, but a distance of less than 5\,kpc, as required to make the \gr\ efficiency similar to that of PSR~B1706$-$44 \citep[${\eta}_{\gamma} = 0.01$, see][]{mal03}, is hard to justify given current observations of the region.  We adopt a nominal distance of 10\,kpc to use in scaling quantities ($d_{10} \equiv d/10$\,kpc) that depend on distance.

\subsection{Neutron Star Thermal Emission and Pulsations}

We find a black-body temperature of $kT_{\infty} = 0.15 \pm 0.02$\,keV for the point source, similar to that found for the Vela pulsar \citep[$kT_{\infty} = 0.128 \pm 0.003$\,keV;][]{pzs+01} and PSR~B1706$-$44 \citep[$kT_{\infty} = 0.14 \pm 0.02$\,keV;][]{ghd02}.  Accounting for the effects of gravitational redshift, this corresponds to a surface temperature $kT_{s} = 0.19 \pm 0.03$\,keV for the canonical neutron-star radius of 10\,km and mass of 1.4\,M$_{\sun}$.  The point source has an unabsorbed flux $F_{\rm X} = 6.3 \times 10^{-13}$ erg cm$^{-2}$ s$^{-1}$ in the range 0.3$-$10.0\,keV.  Scaling the luminosity to a distance of 10\,kpc, $L_{\rm X} = 7.5~{d_{10}}^2 \times 10^{33}$ erg s$^{-1}$ (0.3$-$10.0\,keV).  One can calculate the ratio of the black-body emission radius $R_{e}$ and the distance to the source using the bolometric flux $7.8 \times 10^{-13}$ erg cm$^{-2}$ s$^{-1}$ and the Stefan-Boltzman law.  Correcting for gravitational redshift, this gives $R_{10} / d_{10} = 0.3$, where $R_{10} \equiv R_{e}$/10\,km.  For any reasonable distance, the simple black-body model suggests that the emission is coming from hot spots on the surface, rather than from the entire surface.  This is not unreasonable, as some high-energy emission models predict hot spots due to back-flowing particles.

Using the non-magnetized neutron-star hydrogen and iron atmosphere models of \citet{gbr02}, which assume a neutron-star radius of 10\,km and a mass of 1.4\,M$_{\sun}$, changes the temperature significantly.  Fixing the n$_{\rm H}$ to the value derived from the nebular fit, for a hydrogen atmosphere we get a best-fit value of $kT_{s}$ $ = 0.06 \pm 0.01$\,keV \citep[which is similar to the hydrogen atmosphere fit of the Vela pulsar, $kT_{\infty} = 0.059 \pm 0.003$\,keV;][]{pzs+01} and a flux of $2.9 \times 10^{-14}$ erg cm$^{-2}$ s$^{-1}$ (0.3$-$10.0\,keV).  However, the normalization from this fit implies a $\sim$0.8\,kpc distance, which is unreasonable.  Again fixing the n$_{\rm H}$ to the nebular value, for an iron atmosphere model we get a best-fit value of $kT_{s}$ $ = 0.22 \pm 0.01$\,keV and a flux of $3.6 \times 10^{-14}$ erg cm$^{-2}$ s$^{-1}$ (0.3$-$10.0\,keV).  The implied distance here is $\sim$13\,kpc, which is roughly consistent with the emission coming from the entire surface.

Based on the soft spectrum of the point source, one would expect pulsations to be due primarily to thermal emission.  The profile found by folding the X-ray events of our continuous-clocking data with the radio ephemeris shows a somewhat narrow (duty cycle $\sim$0.3) peak and interpulse, which is possible for thermal pulsations given the right geometry.  If we are seeing emission from hot spots at both poles then it is possible to get two sinusoidal peaks separated by a plateau of emission \citep{bel02}.  In fact, the geometry inferred from the PWN (see \S 4.4) is consistent with the pulse profile seen here.  As noted in \S 2.2, a non-thermal component is also allowed by the spectrum.  The possible pulsations might be a combination of pulsed thermal and magnetospheric emission, but the number of pulsed counts ($\sim$69) is too large to consist solely of magnetospheric emission.  All the known \gr\ pulsars have also been detected as X-ray pulsars with non-thermal emission components to their spectra \citep{krh04}.  We note that a lack of X-ray pulsations would not preclude \psr\ from being a \gr\ pulsar.  For instance, X-ray pulsations in PSR~B1706$-$44 are relatively difficult to detect even though PSR~B1706$-$44 is a well-established \gr\ pulsar.  Thus, \psr\ remains an excellent \gr\ pulsar candidate.  

\subsection{PWN Spectrum}

We found that the spectrum of the nebula is well fit by an absorbed power-law model with n$_{\rm H}=(\errors{7.8}{1.7}{1.4})\times 10^{21}$~cm$^{-2}$ and $\Gamma = \errors{1.7}{0.3}{0.2}$.  The photon index $\Gamma$ is roughly consistent with what is observed from the PWN of other similar pulsars like Vela \citep[$\Gamma = 1.50 \pm 0.04$;][]{} and PSR~B1706$-$44 \citep[$\Gamma = \errors{1.34}{0.24}{0.30}$;][]{ghd02}, although it is somewhat higher.  The relationship of \citet{got03} between the photon index of the nebula and the spin-down energy of the pulsar (equation 3 of that paper) gives $\Gamma = 1.22 \pm 0.33$, which is marginally consistent with the $\Gamma$ measured here.

The unabsorbed nebular flux of \pwn\ is $F_{\rm X} = (\errors{1.9}{0.1}{0.3}) \times 10^{-12}$ erg cm$^{-2}$ s$^{-1}$ in the 0.3$-$10.0\,keV range.  This corresponds to an X-ray luminosity in the same energy range of $L_{\rm{X}} = (2.3~{d_{10}}^2) \times 10^{34}$\, erg s$^{-1}$.  For the same energy range, we calculate an X-ray efficiency ($\eta_{\rm X} \equiv L_{\rm X}/\dot{E}$) of $0.009~{d_{10}}^2$ from the combined point-source and nebular luminosities.  This value fits well with the claimed relationship between the spin-down energy of a pulsar and its nebular X-ray efficiency \citep[c.f.][]{pccm02,che00} even if the distance is as close as 3\,kpc.  However, note that this is only for the bright inner nebula.  If there is significantly more flux coming from diffuse emission (as suggested by the {\it ASCA} data and in analogy with the Vela nebula, see below) then the efficiency will be higher. 

\subsection{PWN Size and Morphology}

The generally accepted view of young PWNe \citep[see][and references therein]{krh04} is that they are synchrotron bubbles blown at the center of an expanding supernova remnant (SNR).  High-energy electrons and positrons, along with magnetic field, are continuously injected into the bubble by the pulsar.  Given that the characteristic age of \psr\ is much larger than the typical time scale for a SNR entering the Sedov phase, it is likely (although not certain since characteristic ages can be off by an order of magnitude or more) that the outer edge of the bubble (at radius $R_{P}$) is expanding subsonically into the shocked supernova ejecta.  The bubble is bounded at the inner edge (at radius $R_T$) by the termination of the ultra-relativistic pulsar wind.

The physical size of \pwn 's inner nebula is $\sim (0.5 \times 1)~d_{10}$\,pc.
As \psr 's spin-down energy is within a factor of two of that of the Vela pulsar, one might naively expect the sizes and morphologies of their nebulae to be comparable.  Scaling the $\sim 10^{\prime \prime}$ radius of \pwn 's inner nebula to that of the Vela pulsar's inner nebula, which has a radius of 52$.^{\prime \prime}$7 \citep{hgh01} and whose distance is well known (290\,pc, Dodson et al. 2003), places \pwn\ at a distance of $\sim$1.5\,kpc, which seems unreasonable given the DM.  In fact, \pwn\ would have to be a few times larger than the Vela nebula to put it at a distance that seems at all reasonable based on its DM.  However, since the Vela inner nebula is only the central part of a much larger and more luminous nebula, a better comparision between the sizes of these two objects could be made if the true extent of \pwn\ is determined in a deeper observation with a larger field of view.

The bright inner nebula of Vela and \pwn\ may be directly downstream of the termination shock of the relativistic pulsar wind.  The location of the termination shock depends primarily on the energy output of the pulsar and the pressure in the nebula.  The pressure in the nebula depends upon the total energy input during the pulsar's lifetime and the losses due to adiabatic and synchrotron processes (possibly complicated by the passage of the SNR reverse shock).  The total energy input is highly dependent on the initial spin period, which is unknown and could range from $\sim$10$-$100\,ms \citep{krv+01,mgb+02}.  We currently do not know the full extent of \pwn , which is likely much larger than the inner nebula seen here.  While it is true that the pressure in the PWN should be balanced by the pressure in the surrounding SNR, we know neither the age (as the characteristic age is a poor estimate of the true age for young pulsars) nor the size (since we have not observed a shell) of the potential remnant.  Therefore, although the Vela pulsar is twice as energetic as \psr , it is entirely possible that \pwn\ is several times larger than the Vela nebula.

The size of the torus places an upper limit on the termination shock radius $R_T$.  In analogy to Eqn. 1 in \citet{rtk+03} \citep[see also][]{che00} we can estimate the post-shock magnetic field to be 

\begin{center}
$B_{N} \sim 2 \times 10^{-5} (\frac{{{\epsilon}_t}^{1/2}}{f_{t}(R_T/r)d_{10}})$\,G 
\end{center}

\noindent where $r$ is the measured location of the torus, ${\epsilon}_t \la 1$ is the ratio of the magnetic energy to the total energy in the post-shock flow and $f_t \la 1$ is a geometrical factor allowing for non-spherical outflow.  Note that the PWN of PSR~B1706$-$44 (which has a virtually identical $\dot{E}$ to \psr ) is both much fainter and smaller than \pwn .  This implies that the magnetic field in PSR~B1706$-$44's nebula is smaller but the density of synchrotron emitting particles is larger.  PSR~B1706$-$44 is a known TeV source \citep{kto+95}, this suggests \psr\ may be an excellent target for ground-based Cerenkov detectors. 

  Our \pwn\ image is rather sparse and Poisson-limited, and so a unique model for its morphology cannot be defined.  Nevertheless, the axisymmetry of \pwn\ suggests that the morphology may be best described as that of a torus, with possibly weak jets directed along the axis of symmetry (see Figure~\ref{pwnfig}).  To investigate the question of \pwn 's morphology quantitatively, we spatially fit the PWN using the relativistic torus fitting described in \citet{nr04}.  Since \pwn\ appears to have a double ridge, our fiducial model is
similar to that of the Vela pulsar PWN: twin tori of radius $r$, symmetrically
spaced by distance $d$ on either side of the pulsar spin equator. The
torus axis is at position angle $\Psi$ (N-E) and inclination $\zeta$ to the 
line of sight. The bulk flow velocity down-stream from the termination
shock is $\beta_T$ and a `blur' parameter $\delta$ is applied to account for
the unmodeled post-shock synchrotron cooling length. A uniform background
and a detailed point source model computed for the pulsar spectrum and
chip position are also included.  The 3-D model is
projected to the plane of the sky, a Poisson-based figure of merit is minimized
to find the best-fit parameters and the statistical errors on these parameters
are estimated by refitting many Poisson realizations of the best-fit model,
each with the same number of counts as seen in the data.

The derived parameters from the fit are summarized in Table~\ref{fit.tab}.  The errors quoted are statistical 1-$\sigma$ values only. Systematic errors
due to unmodeled features are difficult to quantify. One estimate of
their magnitude can be made by excluding an elliptical zone about the pulsar, extended along the symmetry axis where an unmodeled polar jet might be present. Re-fitting
the model to the surrounding data, we obtain values within 1.5~$\sigma$ of
the original fit for all parameters, except $\beta_T$ which now is 
$0.64\pm 0.02$; \citet{nr04} note that for this parameter the 
statistical errors tend to be small and dominated by systematic uncertainties
for edge-on tori.  In this particular case, the value of $\beta_T$ is likely being dominated by the bright clump to the south-west of the pulsar.   We have also fit a single torus; again the geometrical
parameters are quite similar, although a larger blur parameter $\delta
= 1.8^{\prime\prime}$ is required to cover the broader torus. The fit statistic
for the model is of course larger than that of the 2-torus model, by an
amount similar to that corresponding to 1-$\sigma$ excursions in the 
individual fit parameters.
Unfortunately, this does not alone show that the double torus is a
significantly better model; we cannot apply an F-test, since the fit
statistic is not $\chi^2$.

        Clearly a deeper image is needed to draw firm conclusions, but we can
discuss the geometrical implications of the fit angles. The large $\zeta$
implies that we are viewing the pulsar near the spin equator. In the outer
magnetosphere picture of \citet{ry95} this is indeed where
we expect to see $\gamma$-ray emission. More particularly, Figure 4 of that
paper shows that for $\zeta \approx 83^\circ$ and magnetic impact angle $\beta$
modest (to allow detection of the radio emission), we expect two $\gamma$-ray
pulses separated by $\sim$180$^\circ$, with the first lagging the radio peak
by $\sim$10$^\circ$. Intriguingly, \citet{mc03} found apparently
significant pulsations folding the $\gamma$-ray photons from this
source near the extrapolated
$P$ for two viewing periods; the light curves show two narrow peaks separated
by 0.5 in phase.  It is important to note that for $\zeta$ near 90$^\circ$,
such a light curve is also naturally expected in the two pole model of
\citet{dr03} and may be possible in polar-cap models \citep{dh96}.
The separation of the two tori suggest bright emission at spin axis angles
$\theta_T = {\rm tan^{-1}} (2r {\rm sin}\zeta/d) = 77^\circ, 103^\circ$,
as viewed from the pulsar. If these represent enhanced e$^\pm$ flowing radially
from near the magnetic poles then we might infer a magnetic inclination 
$\alpha \approx 77^\circ$.
This would imply an impact parameter $\beta=\zeta-\alpha \approx  +6^\circ$.  Polarization sweep measurements could test this picture.
The unscattered pulse half-power width $\sim$33$^\circ$ of \psr\ is
substantially wider than the 
$2.5^\circ (P/1\,{\rm s})^{-1/2}/{\rm sin}\alpha$ width expected
for a core component, but is similar to the 
$5.8^\circ (P/1\,{\rm s})^{-1/2}$ width for an outer cone \citep{ran93}.
Of all the fit geometrical parameters, the most robust is the fit
PA of the symmetry axis $\Psi$. Although difficult to measure, given the 
large DM and expected substantial Faraday rotation,
the absolute position angle at the polarization sweep
maximum would provide an interesting check of the inference that this
represents the pulsar spin axis.

\section{Conclusions and Further Work}

We have used the {\it Chandra X-ray Observatory} to observe the young and energetic pulsar \psr\ and have detected a new PWN, \pwn , as well as an embedded point source near the radio timing position of the pulsar.  The morphology of \pwn\ is perhaps best decribed as an equatorial torus, and we have derived geometrical parameters for the orientation of the system by spatially fitting the nebula.  Spectral fitting of the nebula shows that it is similar to the nebulae of other energetically-similar pulsars.  Spectral fitting of the point source shows that it is well fit by an absorbed black-body model or a black-body plus power-law model.  Radio timing observations of \psr\ have revealed a glitch similar to the largest glitches seen in the Vela pulsar.
Analysis of countinuous-clocking data from {\it Chandra} indicates that there may be weak X-ray pulsations from the source (pulsed fraction of $\sim$37\%) at the period predicted by the radio ephemeris.  \psr\ remains the most likely counterpart to the high-energy \EG\ \gr\ source GeV J2020+3651 (which is the 10th brightest \gr\ source above 1\,GeV).  It is an excellent \gr\ pulsar candidate for the AGILE and GLAST missions, although a contemporaneous ephemeris will be essential, given the large amount of timing noise the pulsar exhibits.  

Deeper observations with {\it Chandra} and {\it XMM-Newton} are needed to clarify the morphology of \pwn\ and to confirm the presence of pulsations.  Such observations are also able to confirm whether the faint outer jet hinted at here is real.  Currently, there only about a half dozen PWNe with torus plus jet morphologies, with the Crab and (likely) Vela nebulae as the best examples.  With a deeper observation, \pwn\ may prove to be the third best example.   Further work by our group will also include analysis of radio polarization data we have taken with Arecibo.  These data may help bolster the geometrical interpretation described here if the swing of the polarization angle across the pulse is measurable.  Proper-motion measurements could check if \psr\ is moving along the axis of the nebula, which is the case for both Vela and the Crab.  However, measuring the proper motion of \psr\ (which will likely be small since the distance is likely large) will be very difficult both through radio timing because of timing noise or through VLBI because the radio source is very faint.  We are also analyzing VLA observations (in A, C, and D arrays) of the source region to look for a radio PWN and/or SNR.  This is important for characterizing the surrounding medium, which has an important effect on the morphology of the PWN and its size.

\acknowledgements

J.W.T.H. is an NSERC PGS A fellow.  
V.M.K. gratefully acknowledges support from NSERC Discovery Grant
228738-03, NSERC Steacie Supplement 268264-03, a Canada Foundation for
Innovation New Opportunities Grant, FQRNT Team and Centre Grants, and NASA
Long-Term Space Astrophysics Grant NAG5-8063.  V.M.K. is a Canada Research
Chair and Steacie Fellow.
R.W.R. acknowledges support from NASA grants NAGS-13344 
and SAOG03-4093B.  The Arecibo Observatory is part of the National Astronomy 
and Ionosphere Center, which is operated by Cornell University under a 
cooperative agreement with the National Science Foundation. 

\bibliographystyle{apj}


\begin{deluxetable}{l l}
\tablewidth{0pt}
\tablecaption{Measured and Derived Parameters for PSR~J2021+3651 \label{timing.tab}}
\tablehead{\colhead{Parameter} & \colhead{Value}}  

\startdata

Right ascension (J2000.0)\tablenotemark{a}    &  $20^h21^m05^s.46(5)$  \\
Declination (J2000.0)\tablenotemark{a}    & $+36^{\circ}51^{\prime}04^{\prime \prime}.8(7)$ \\
Galactic longitude $l$ (deg)\tablenotemark{a} &  75.23        \\
Galactic latitude  $b$ (deg)\tablenotemark{a} &  +0.11	      \\
DM  (pc cm$^{-3}$)   			      &  369.2(2)      \\

\multicolumn{2}{c}{Preglitch} 				      \\
Pulse frequency $\nu$ (s$^{-1}$)              &  9.64113518458(16)       \\
Frequency derivative $\dot{\nu}$ (s$^{-2}$)   & $-8.886500(24)\times 10^{-12}$ \\
Epoch (MJD)  				      &  52407.389     \\
Range of MJDs  				      &  52305$-$52616    \\	
RMS Residuals (ms) 			      &  3.0     \\

\multicolumn{2}{c}{Postglitch} \\
Pulse frequency $\nu$ (s$^{-1}$)              &  9.64091197572(12)       \\
Frequency derivative $\dot{\nu}$ (s$^{-2}$)   & $-8.932592(12)\times 10^{-12}$     \\
Epoch (MJD)   				      &   52730     \\
Range of MJDs   			      &  52645$-$53009    \\
RMS Residuals (ms)	  		      &  9.3      \\

\multicolumn{2}{c}{Glitch Parameters} \\
Epoch (MJD) 				      & 52629.97$-$52630.16 \\
\vspace{1mm}
Frequency jump ${\Delta \nu}/{\nu}$           & $2.587(2)\times 10^{-6}$  \\
\vspace{1mm}
Frequency derivative jump ${\Delta \dot{\nu}}/{\dot{\nu}}$   & $6.2(3)\times 10^{-3}$   \\

\multicolumn{2}{c}{Derived Parameters} \\
Spin-down Luminosity $\dot{E}$\tablenotemark{b} ~(ergs s$^{-1}$)    & $3.4\times 10^{36}$  \\
Surface Magnetic Field $B$\tablenotemark{c} ~(G)                    & $3.2 \times 10^{12}$  \\
Characteristic Age ${\tau}_c \equiv P/2\dot{P}$ (kyr)   &  17 \\
\enddata
\tablecomments{ Figures in parentheses represent uncertainty in the least-significant digits quoted equal to 3 times errors given by {\tt TEMPO}.}
\tablenotetext{a}{ Coordinates of the {\it Chandra} X-ray point source.  We estimate a positional error of $\sim$1$^{\prime \prime}$ by comparing the position of another bright X-ray point source on the S3 chip with a catalogued optical counterpart.} 
\tablenotetext{b}{ $\dot E \equiv 4 {\pi}^2 I \dot P / P^3$ with $I = 10^{45}$ g cm$^2$.}
\tablenotetext{c}{ Assuming standard magnetic dipole spindown:
$B \equiv 3.2 \times 10^{19} (P \dot P)^{1/2}$ Gauss.}

\end{deluxetable}

\begin{deluxetable}{l l}
\tablewidth{0pt}
\tablecaption{Torus Fit Parameters \label{fit.tab}}
\tablehead{\colhead{Parameter\tablenotemark{a}} & \colhead{Value}}  

\startdata

$\Psi$     &    $47^{\circ} \pm 1^\circ$ \\
$\zeta$    &    $83^{\circ} \pm 1^\circ$ \\
$r$        &    $8.6^{\prime\prime} \pm 0.3^{\prime\prime}$ \\
$\delta$\tablenotemark{b}   &    $1.2^{\prime\prime}$ \\
$\beta_T$\tablenotemark{c}  &    $\sim$0.8 \\
$d$        &    $3.7^{\prime\prime} \pm 0.2^{\prime\prime}$ \\

\enddata

\tablenotetext{a}{Explanation of parameters in \S 4.4}

\tablenotetext{b}{Held fixed in the global fit.}

\tablenotetext{c}{The Doppler boosting is likely being dominated
by the bright clump to the south-west of the pulsar.}

\end{deluxetable}


\begin{figure}
\plotone{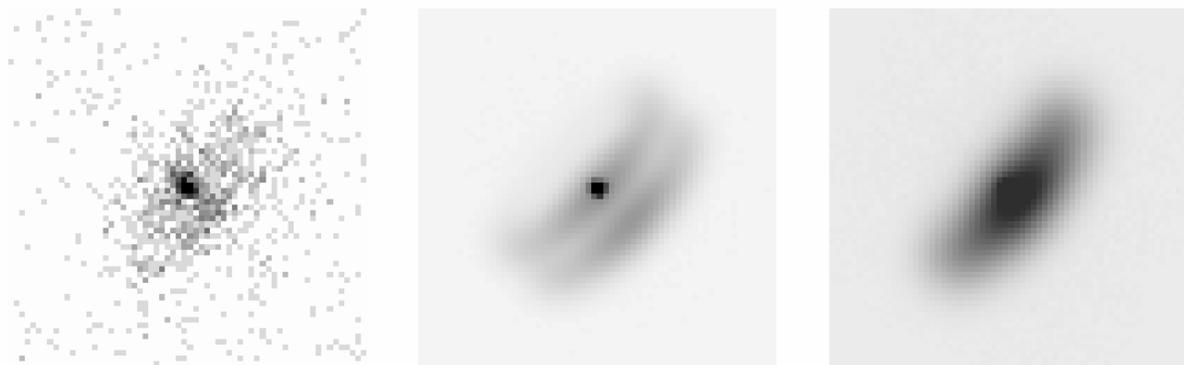}
\caption{{\it Left}: 32$^{\prime \prime} \times 32^{\prime \prime}$ 
cutout of the central PWN.
{\it Middle}: best-fit
double torus plus point source model. {\it Right}: best-fit single torus
plus point source model.
\label{pwnfit}
}

\end{figure}

\end{document}